\documentclass[
aps, 
pra, 
reprint, 
floatfix, 
footinbib, 
longbibliography, 
]{revtex4-1}

\usepackage{amsmath} 
\usepackage{amsfonts} 
\usepackage{amssymb} 
\usepackage{amsthm} 
\usepackage{mathtools} 
\usepackage{braket} 
\usepackage{graphicx} 
\usepackage{enumerate} 
\usepackage{tabularx} 
\usepackage[breaklinks]{hyperref} 
\usepackage[dvipsnames]{xcolor} 

\theoremstyle{definition}

\newtheorem*{theorem*}{Security of the MPQC protocol}

\newtheorem*{definition*}{Definition of the $\epsilon$-security}

\newtheorem*{lemma*}{Lemma}

\newtheorem*{property*}{Property}

\begin{document}

\title{An improvement on the versatility of secure multi-party quantum computation protocol: exploitation of triorthogonal quantum
error-correcting codes}

\author{Petr A. Mishchenko}
\email{petr.mishchenko.us@hco.ntt.co.jp}

\author{Keita Xagawa}
\email{keita.xagawa.zv@hco.ntt.co.jp}

\affiliation{NTT Social Informatics Laboratories, Tokyo 180-8585, Japan}

\date{\today}


\begin{abstract}
Secure multi-party quantum computation (MPQC) protocol is a versatile tool that enables error-free distributed quantum computation to a
group of $n$ mutually distrustful quantum nodes even when some of the quantum nodes do not follow the instructions of the protocol
honestly. However, in case of the MPQC protocols built on top of the quantum error correction technique, the versatility is significantly
affected by the fact that one has to choose a particular quantum error-correcting code (QECC), which immediately applies a constraint on
the number of quantum nodes $n$. Therefore, in this talk, we suggest a modified MPQC protocol based on triorthogonal QECCs which applies
significantly less constraint on the number of quantum nodes $n$ if compared to the previously suggested MPQC protocol based on triply-even
QECCs. Especially, the variety of available options in the region of a small number of quantum nodes $n$ becomes important in the noisy
intermediate-scale quantum (NISQ) era.
\end{abstract}


\maketitle

\section{
  \label{sec:introduction}
  Introduction
}

Secure multi-party quantum computation (MPQC) protocol is a tool that allows $n$ quantum nodes to jointly compute some publicly known
quantum circuit on their private inputs in a distributed manner~\cite{crepeau_stoc_02_643-652_2002,smith_arxiv:quant-ph/0111030}. In more
detail, MPQC protocol can be described as a cryptographic primitive where each quantum node $i$ inputs some quantum state $\rho_i$ and then
$n$ quantum nodes jointly perform arbitrary quantum circuit $\mathcal{U}$ with $n$ inputs and $n$ outputs. Finally, each quantum node $i$
gets some output quantum state $\omega_i$. During the execution of the MPQC protocol $t$ cheating quantum nodes cannot affect output of the
computation beyond choosing their own inputs as well as cannot obtain any information on the inputs of the honest quantum nodes beyond what
they can infer from the output of the computation.

In a nutshell, MPQC protocols developed for the quantum circuits can be divided into two types: information-theoretically secure ones based
on a technique of quantum error correction which are able to tolerate $t < \frac{n}{4}$ cheating quantum
nodes~\cite{crepeau_stoc_02_643-652_2002,smith_arxiv:quant-ph/0111030,lipinska_pra_102_022405_2020,mishchenko_arxiv:2206.04871}, and
computationally secure ones based on a technique of quantum authentication codes which are able to tolerate $t < n$ cheating quantum
nodes~\cite{dulek_eurocrypt_12107_729-758_2020,alon_crypto_12825_436-466_2021}.

In this talk, we focus on the former type of the MPQC protocols and utilize verifiable hybrid secret sharing (VHSS) protocol suggested in
Ref.~\cite{lipinska_pra_101_032332_2020}. The VHSS protocol rests on the Steane-type quantum error correction
technique~\cite{steane_prl_78_2252_1997} and works for any type of the Calderbank-Shor-Steane (CSS) quantum error-correcting codes
(QECCs)~\cite{steane_prsla_452_2551-2577_1996,calderbank_pra_54_1098_1996} encoding single-qubit quantum states into $n$-qubit logical
quantum states.

In the MPQC protocols utilizing the VHSS protocol, see Refs.~\cite{lipinska_pra_102_022405_2020,mishchenko_arxiv:2206.04871}, each quantum
node $i$ encodes and shares his single-qubit input quantum state $\rho^i$. In such a way, quantum nodes create global logical quantum state
shared among all the $n$ quantum nodes, and as a consequence, each quantum node $i$ holds a part of the global logical quantum state called
a \textit{share}. Then, quantum nodes jointly verify the encoding of each single-qubit quantum state $\rho^i$ using the VHSS protocol.
After that, quantum nodes locally perform quantum operations on their \textit{shares} in order to evaluate the logical version of the
quantum circuit $\mathcal{U}$. Finally, each quantum node $i$ collects all the \textit{shares} corresponding to his output and by decoding
these \textit{shares} reconstructs his single-qubit output quantum state $\omega_i$.

Indeed, it would be preferable to implement the MPQC protocols described above with as few qubits as possible since in the noisy
intermediate-scale quantum (NISQ) era physical devices will be severely limited in the number of available qubits. The first step towards
reduction in the number of required qubits from the initial suggestion in
Refs.~\cite{crepeau_stoc_02_643-652_2002,smith_arxiv:quant-ph/0111030} was made in Ref.~\cite{lipinska_pra_102_022405_2020} where the number
of necessary qubits per quantum node was reduced from $\Omega \left( (n^3 + n^2 s^2) \log n \right)$ to $n^2 + \Theta(s) n$ by taking
advantage of the self-dual CSS QECCs~\cite{steane_prsla_452_2551-2577_1996,calderbank_pra_54_1098_1996} encoding single-qubit quantum
states into $n$-qubit logical quantum states. After that, by taking advantage of the triply-even CSS
QECCs~\cite{betsumiya_jlms_86_1_1-16_2012,knill_arxiv:quant-ph/9610011}, also encoding single-qubit quantum states into $n$-qubit logical
quantum states, the number of necessary qubits per quantum node was further reduced from $n^2 + \Theta(s) n$ to $n^2 + 3n$ in
Ref.~\cite{mishchenko_arxiv:2206.04871}. However, tirply-even CSS QECCs has a significant drawback in terms of constraint on the allowed
number of quantum nodes $n$, especially in the region of small $n$. We are aware only of two triply-even CSS QECCs with the requirements
necessary for the implementation of the MPQC protocol: $[[n = 15, k = 1, d = 3]]$ QECC~\cite{knill_arxiv:quant-ph/9610011} and
$[[n = 49, k = 1, d = 5]]$ QECC~\cite{bravyi_pra_86_052329_2012}.

In order to achieve less constraint on the number of allowed quantum nodes $n$ we suggest MPQC protocol constructed on the basis of
triorthogonal CSS QECCs~\cite{bravyi_pra_86_052329_2012} which actually constitute a superclass for triply-even CSS
QECCs~\cite{betsumiya_jlms_86_1_1-16_2012,knill_arxiv:quant-ph/9610011}. Especailly, we were motivated by the recent classification of
triorthogonal CSS QECCs in Ref.~\cite{nezami_pra_106_012437_2022}, which shows a variety of available options for the number of quantum
nodes in the region of small $n$. For example, one can find $[[n = 23, k = 1, d = 3]]$ QECC or $[[n = 27 \sim 37, k = 1, d = 3]]$ QECCs.
Therefore, exploitation of triorthogonal CSS QECCs largely enhances versatility of the MPQC protocol and makes it more accessible in the
NISQ era.

\section{
  \label{sec:assumptions_and_definitions}
  Assumptions and Definitions
}

\textbf{Triorthogonal CSS QECCs:} Here we briefly explain how to construct stabilizer codes from triorthogonal
matrices~\cite{bravyi_pra_86_052329_2012}. Suppose there exist two binary vectors $f$, $g \in \{0, 1\}^n$ with hamming weights $|f|$ and
$|g|$ respectively, and have entry-wise product $f \cdot g \in \{0, 1\}^n$. We call an $m \times n$ binary matrix $G$ triorthogonal if for
its rows $f_1, \ldots, f_m \in \{0, 1\}^n$ following two conditions are satisfied for all pairs $i$, $j$ and triples $i$, $j$, $k$ with
distinct indices.
\begin{align}
&|f_i \cdot f_j| = 0 \pmod 2 \\
&|f_i \cdot f_j \cdot f_k| = 0 \pmod 2
\end{align}

From an $m \times n$ binary matrix $G$ defined in this way one can construct an $n$-qubit triorthogonal CSS QECC as
follows~\cite{bravyi_pra_86_052329_2012}. For each even-weight row of $m \times n$ binary matrix $G$ one defines an $X$ stabilizer
generator by mapping non-zero entries to $X$ operators. Next, for each even-weight row of the orthogonal complement of the binary matrix
$G$, i.e., $G^\bot$, one defines a $Z$ stabilized generator by mapping non-zero entries to $Z$ operators.

In case of triorthogonal CSS QECCs one may decide on the universal set of quantum gates composed of $CCZ$ and $H$
gates~\cite{shi_qic_3_84_2003,aharonov_arXiv:quant-ph/0301040}, among which $CCZ$ gate can be implemented
transversally~\cite{paetznick_prl_111_090505_2003}. On the other hand, $H$ gate cannot be implemented transversally and therefore needs to
be implemented by the gate teleportation technique~\cite{gottesman_nature_402_390-393_1999} in a similar manner to
Refs~\cite{lipinska_pra_102_022405_2020,mishchenko_arxiv:2206.04871}.

\textbf{Communication channels:} We assume that all the quantum nodes have an access to the classical authenticated broadcast
channel~\cite{canetti_ieee_99_2_708-716_1999} and to the public source of randomness~\cite{rabin_stoc_89_73-85_1989}. Also, each pair of
quantum nodes is connected via the authenticated and private classical~\cite{canetti_ieee_17_219-233_2004} and
quantum~\cite{barum_ieee_43_449-458_2002} channels.

\textbf{Adversary:} We make no assumptions on the computational power of the adversary. Our non-adaptive active adversary is limited only
by the number of quantum nodes $t$ that it can corrupt. These corrupted quantum nodes are the cheating quantum nodes mentioned above.

\section{
  \label{sec:mpqc_summary}
  Summary of the MPQC protocol
}

\begin{table}[htb]
  \caption{
    Summary of the MPQC protocol.
  }
  \begin{ruledtabular}
    \begin{tabularx}{\linewidth}{X}
      \textbf{Input:} Single-qubit quantum state $\rho^i$ from each quantum node $i$, agreement on a particular triorthogonal CSS QECC, and
      on a particular quantum circuit $\mathcal{U}$. \\ \\
      \textbf{Output:} In case of success, single-qubit quantum state $\omega_i$ in the possession of each quantum node $i$. In case of
      failure, i.e., excess in the number of cheating quantum nodes, the MPQC protocol is aborted at the end of execution. \\
      \begin{enumerate}
        \item \textbf{Sharing:} By encoding and sharing each of the single-qubit quantum states $\rho^1, \ldots, \rho^n$ \textit{twice},
        quantum nodes create global logical quantum state $\bar{\bar{\mathcal{P}}}$ where each quantum node $i$ holds a share
        $\bar{\bar{\mathcal{P}}}_i$.
        \item \textbf{Verification:} All the quantum nodes jointly verify the encoding of each single-qubit quantum state $\rho^i$ with the
        help of the VHSS protocol in order to check whether each quantum node $i$ is honest or not.
        \item \textbf{Computation:}
        \begin{enumerate}
          \item If for the implementation of the logical quantum circuit $\bar{\bar{\mathcal{U}}}$ the ancillary logical quantum state
          $\bar{\bar{\ket{0}}}^i$ is required, quantum nodes jointly prepare it by the VHSS protocol.
          \item Depending on whether the quantum gate appearing in the logical quantum circuit $\bar{\bar{\mathcal{U}}}$ can be implemented
          transversally or not, quantum nodes behave in the following two ways:
          \begin{enumerate}
            \item In case of the transversal $CCZ$ gate, each quantum node $i$ locally applies corresponding quantum operations to his hare
            $\bar{\bar{\mathcal{P}}}_i$.
            \item In case of the non-transversal $H^i$ gate, quantum nodes jointly prepare verified by the VHSS protocol ancillary logical
            quantum state $\bar{\bar{\ket{+}}}^i$, and then perform the gate teleportation protocol, during which information on the
            positions of the cheating quantum nodes is updated.
          \end{enumerate}
        \end{enumerate}
        \item \textbf{Reconstruction:} Each quantum node $i$ collects all the single-qubit quantum states corresponding to his output and
        by decoding in such a way obtained output logical quantum state $\bar{\bar{\Omega}}^i$ \textit{twice} reconstructs his single-qubit
        output quantum state $\omega_i$.
      \end{enumerate}
    \end{tabularx}
  \end{ruledtabular}
  \label{table:mpqc_summary}
\end{table}

Here we briefly describe our MPQC protocol, see Table~\ref{table:mpqc_summary}. First of all, quantum nodes agree on a particular
triorthogonal CSS QECC~\cite{bravyi_pra_86_052329_2012} encoding single-qubit quantum states into $n$-qubit logical quantum states. After
that, quantum nodes create global logical quantum state $\bar{\bar{\mathcal{P}}}$ (double bar means that the initial input quantum state
from $n$ quantum nodes $\mathcal{P}$ is encoded \textit{twice}) by encoding and sharing each of the single-qubit input quantum states
$\rho^1, \ldots, \rho^n$ \textit{twice}. As a result, each quantum node $i$ holds a part of the global logical quantum state
$\bar{\bar{\mathcal{P}}}$, i.e., a share denoted as $\bar{\bar{\mathcal{P}}}_i$. Next, in order to check whether each quantum node $i$ is
honest or not, quantum nodes jointly verify the encoding of each single-qubit quantum state $\rho^i$ by using the VHSS
protocol~\cite{lipinska_pra_101_032332_2020}. After that, quantum nodes jointly evaluate logical quantum circuit $\bar{\bar{\mathcal{U}}}$.
Here, in case of the transversal $CCZ$ gates, each quantum node $i$ locally performs necessary quantum operations on his share
$\bar{\bar{\mathcal{P}}}_i$. On the other hand, in case of non-transversal $H$ gates, quantum nodes jointly perform the gate teleportation
protocol as will be explained below. Moreover, if implementation of the logical quantum circuit $\bar{\bar{\mathcal{U}}}$ requires
ancillary logical quantum state $\bar{\bar{\ket{0}}}^i$, quantum nodes jointly prepare it by the VHSS protocol. Finally, each quantum node
$i$ collects all the quantum states corresponding to his output and by decoding in such a way obtained output logical quantum state
$\bar{\bar{\Omega}}^i$ \textit{twice} eventually reconstructs his single-qubit output quantum state $\omega^i$. Also, during the execution
of the MPQC protocol quantum nodes publicly record the positions of the cheating quantum nodes in order to decide whether to abort the MPQC
protocol or not.

\begin{figure}[htb]
  \begin{center}
    \includegraphics[scale=0.86]{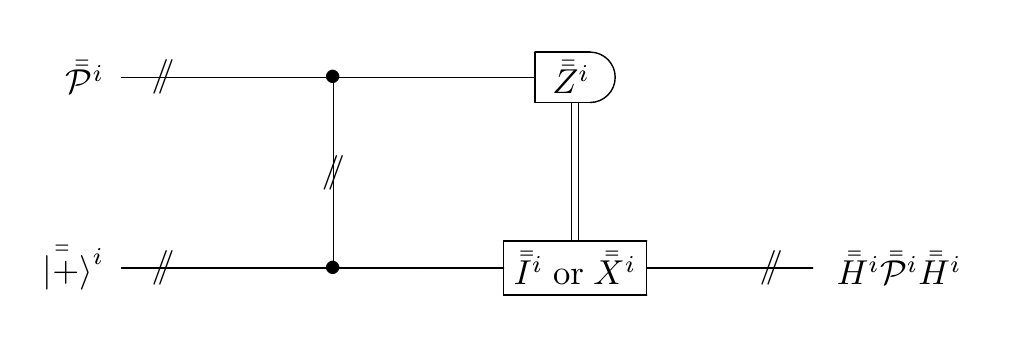}
    \caption{
      Fragment of the logical quantum circuit $\bar{\bar{\mathcal{U}}}$ in which quantum nodes apply a non-transversal $\bar{\bar{H}}^i$
      gate to the logical quantum state $\bar{\bar{\mathcal{P}}}^i$ with the gate teleportation technique.
    }
    \label{fig:gate_teleportation_h}
  \end{center}
\end{figure}

In short, a gate teleportation technique implementing a non-transversal $H^i$ gate (superscript $i$ means that the quantum gate is applied
to the quantum wire $i$ of the quantum circuit $\mathcal{U}$) is performed as follows. Suppose quantum nodes want to apply a
non-transversal $\bar{\bar{H}}^i$ gate, applied to the part of the global logical quantum state $\bar{\bar{\mathcal{P}}}$ initially created
from the single-qubit quantum state $\rho^i$ and denoted as $\bar{\bar{\mathcal{P}}}^i$. Quantum nodes jointly prepare verified by the VHSS
protocol ancillary logical quantum state $\bar{\bar{\ket{+}}}^i$. Then, with the help of logical quantum gates ($CZ$ and $X$ gates can be
implemented transversally), logical measurements (which also can be implemented transversally) and classical communication, quantum nodes
apply non-transversal $\bar{\bar{H}}^i$ gate to the logical quantum state $\bar{\bar{\mathcal{P}}}^i$, see
Fig.~\ref{fig:gate_teleportation_h}. During the gate teleportation protocol information on the positions of the cheating quantum nodes is
updated.

We note that our MPQC protocol has an exponentially small probability of error inherited from the VHSS protocol, i.e.,
$\kappa 2^{-\Omega(r)}$, where $r$ is the security parameter and $\kappa$ is standing for the number of times the VHSS protocol is invoked.
The number of cheating quantum nodes the MPQC protocol can tolerate is constrained by the distance $d$ of the triorthogonal CSS QECC as
$t \leq \left \lfloor \frac{d - 1}{2} \right \rfloor$. This constraint allows honest quantum nodes to correct all the arbitrary quantum
errors introduced by the $t < \frac{n}{4}$ cheating quantum nodes. Important to note that we allow our MPQC protocol to abort at the end of
execution if honest quantum nodes detect too many cheating quantum nodes during the execution of the protocol, in a similar manner to
Refs.~\cite{lipinska_pra_102_022405_2020,mishchenko_arxiv:2206.04871}.

During the execution of the MPQC protocol, in addition to the $n^2$ single-qubit quantum states required for holding a share
$\bar{\bar{\mathcal{P}}}_i$, each quantum node $i$ uses $2n$ single-qubit ancillary quantum states to verify the encodings of single-qubit
input quantum states $\rho^1, \ldots, \rho^n$ with the VHSS protocol, and $3n$ single-qubit ancillary quantum states to apply a
non-transversal $H^i$ gate with the gate teleportation protocol involving verification of the ancillary logical quantum state
$\bar{\bar{\ket{+}}}^i$, or to verify the ancillary logical quantum states $\bar{\bar{\ket{0}}}^i$ which may be required for the
implementation of the logical quantum circuit $\bar{\bar{\mathcal{U}}}$. Thus, in total each quantum node requires $n^2 + 3n$ qubits for
the implementation of the MPQC protocol.

\section{
  \label{sec:summary}
  Summary
}

To summarize, we suggested MPQC protocol built upon a technique of quantum error correction and in particular constructed on the basis of
triorthogonal CSS QECCs which applies significantly less constraint on the number of quantum nodes $n$ if compared to the previously
suggested MPQC protocol constructed on the basis of triply-even CSS QECCs. In essence, triorthogonal CSS QECCs constitute a superclass for
triply-even CSS QECCs. In addition, in case of triply-even CSS QECCs we are aware only of two triply-even CSS QECCs with the requirements
necessary for the implementation of the MPQC protocol: $[[n = 15, k = 1, d = 3]]$ QECC and $[[n = 49, k = 1, d = 5]]$ QECC, while in case
of triorthogonal CSS QECCs one can find a variety of available options: $[[n = 23, k = 1, d = 3]]$ QECC or
$[[n = 27 \sim 37, k = 1, d = 3]]$ QECCs. Such a diversity of available options, especially in the region of small $n$, largely enhances
versatility of the MPQC protocol and makes it more accessible in the NISQ era.


\bibliography{main_0}


\end{document}